\documentclass[conference]{IEEEtran}
\IEEEoverridecommandlockouts
\usepackage{tabularx}
\usepackage[T1]{fontenc}
\usepackage[noend]{algpseudocode}
\usepackage{graphicx}
\usepackage{amsmath,amssymb,amsfonts,stfloats}
\usepackage{array}
\usepackage{textcomp}
\usepackage[font=small]{caption}
\usepackage{paralist}
\usepackage{lettrine}
\usepackage{cite}
\usepackage{xcolor}
\usepackage{cancel}
\usepackage{booktabs}
\usepackage{mathtools}
\usepackage{array}
\usepackage{booktabs}
\usepackage[caption=false,font=footnotesize]{subfig}
\usepackage{url}
\usepackage{acronym}
\usepackage{multicol}
\usepackage{makecell}
\newcommand{\norm}[1]{\left\lVert#1\right\rVert}

\begin{document}

\title{Advanced Tri-Sectoral Multi-User Millimeter-Wave Smart Repeater}  
\author{Kai Dong, Silvia Mura, Marouan Mizmizi, Dario Tagliaferri, and Umberto Spagnolini \\
\IEEEauthorblockA{Dipartimento di Elettronica, Informazione e Bioingegneria (DEIB), Politecnico di Milano, Milan, Italy. \\
Email: \{kai.dong, silvia.mura, marouan.mizmizi, dario.tagliaferri, umberto.spagnolini\}@polimi.it}
}
\maketitle

\begin{abstract}
Smart Repeaters (SR) can potentially enhance the coverage in Millimeter-wave (mmWave) wireless communications. However, the angular coverage of the existing two-panel SR is too limited to make the SR a truly cost-effective mmWave range extender. This paper proposes the usage of a tri-sectoral Advanced SR (ASR) to extend the angular coverage with respect to conventional SR. We propose a multi-user precoder optimization for ASR in a downlink multi-carrier communication system to maximize the number of served User Equipments (UEs) while guaranteeing constraints on per-UE rate and time-frequency resources.  
Numerical results show the benefits of the ASR against conventional SR in terms of both cumulative spectral efficiency and number of served UEs (both improved by an average factor 2), varying the system parameters.
\end{abstract}

\section{Introduction}
The 5G network rollouts are in full swing globally, with standardization advancing to address new market verticals such as automotive, energy, and industrial internet of things \cite{3GPPTR38901}. One of the main innovations of 5G and the primary feature of future 6G networks is the high frequency, e.g., mmWave ($30$ - $100$ GHz), sub-THz, and THz bands \cite{asghar2022evolution}. Radio propagation at these frequencies is subject to a strong path attenuation and is severely affected by link blockage \cite{rappaport2019wireless}. Hence, communication is efficient only for a limited range. Thereby, a denser deployment of network infrastructures is required. However, exclusively employing Base Stations (BSs) represents a questionable strategy with prohibitive deployment and maintenance costs.
The recently introduced Reconfigurable Intelligent Surfaces (RISs) represent a low-cost solution to complement the network infrastructure, extending its coverage \cite{zeng2020reconfigurable} and bypassing blocked links \cite{kishk2020exploiting}. However, RISs are not fully mature for large-scale deployment, as several technological challenges need to be addressed, e.g., the real-time configuration in high mobility \cite{cao2021ai}. 

%

\begin{figure}[!tb]  
 	\centering
	\subfloat[Conventional two-panel SR]{\includegraphics[width=0.95\columnwidth]{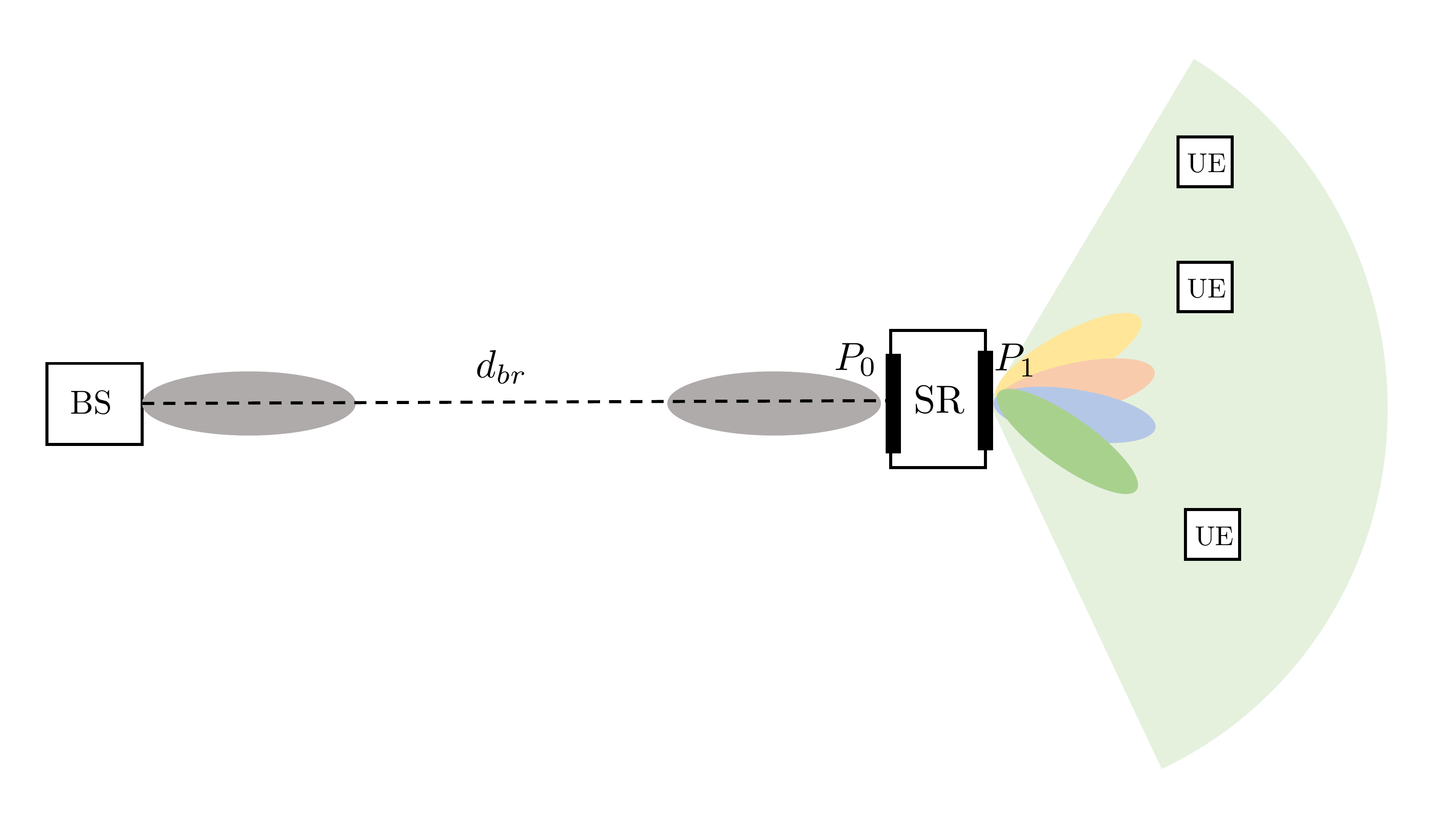} \label{subfig:SR}} \\
    \subfloat[Advanced tri-panel SR] {\includegraphics[width=0.95\columnwidth]{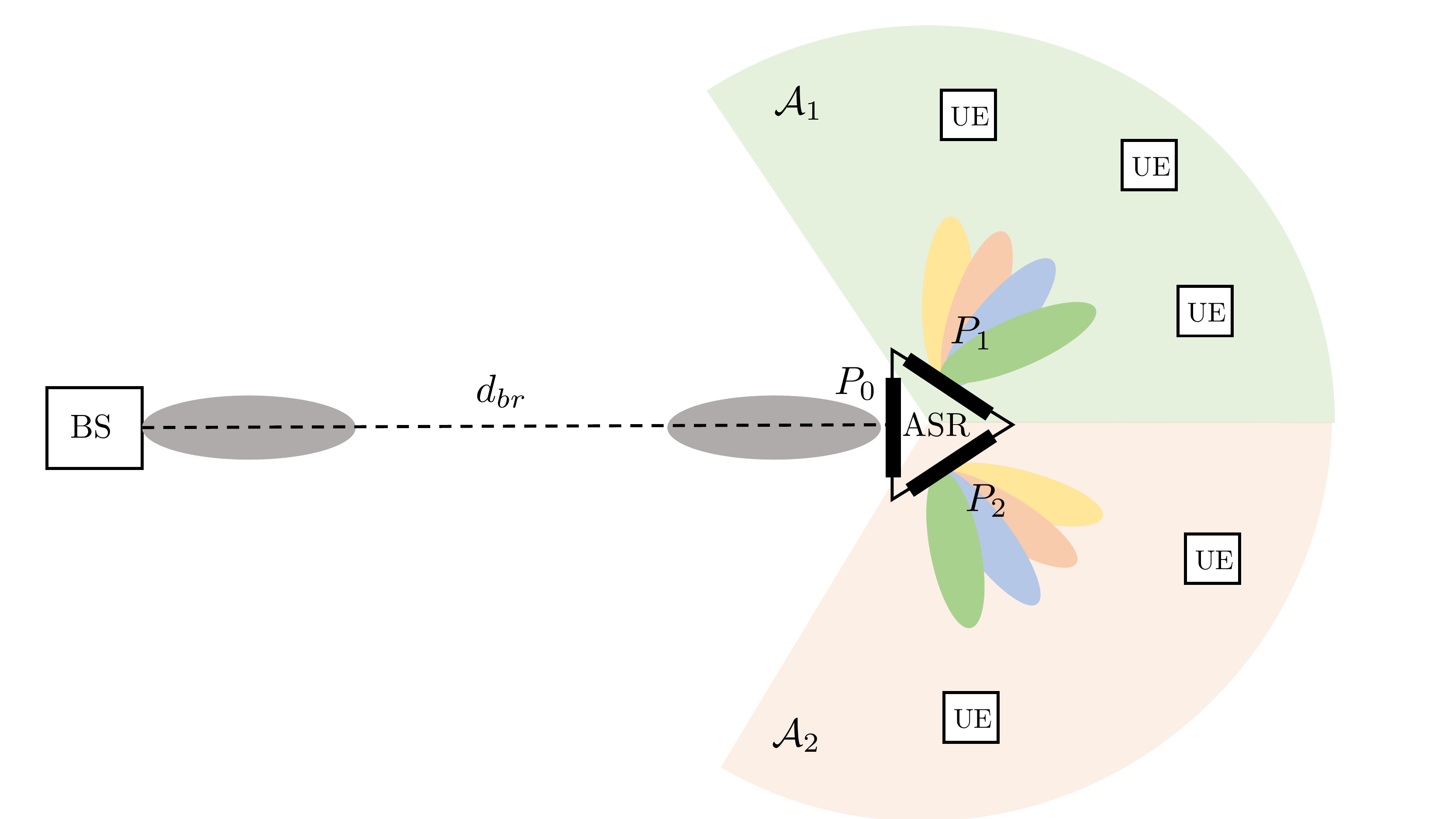}\label{subfig:ASR}}
    \caption{Top view of the BS-UE downlink scenario assisted by: (a) conventional SR; (b) proposed ASR. The conventional SR covers an angular sector of 120 deg, while the ASR coverage is doubled. The remaining 120 deg is covered by the BS.}
    \label{fig:FrameworkAFRelay}
\end{figure}

Alternatively, Smart Repeaters (SR), a.k.a. network-controlled repeaters, are being considered by the Third Generation Partnership Project (3GPP) in Release 18 for 5G networks deployment to extend the coverage provided by the BSs \cite{SRZTE2021}. The SRs can be regarded as the evolution of the classical RF repeaters based on Amplify-and-Forward (AF) operation, which is low-cost and easy to deploy \cite{Flamini9720231}. The SRs use side information to achieve a more intelligent AF operation in a system with Time-Division Duplex (TDD) access and beamforming operation \cite{SRQua2021}. From the hardware perspective, SR consists of two beamforming antennas (e.g., phased antenna arrays), one oriented toward the serving BS and the other toward the service area to be covered, as illustrated in Fig. \ref{subfig:SR}.

The key challenge addressed in this paper is the design of the precoding/combining at the SR to enable multi-user functionality while extending the network coverage. 
The recent literature focuses on hybrid Multiple-Input Multiple-Output (MIMO) architectures with several RF chains, which enable high communication efficiency while increasing the deployment costs and the system complexity and introducing a delay due to base-band processing. The authors in \cite{jiang2019mmwave} proposed a single-user multi-stream precoder and combiner in an AF relay based on a hybrid antenna array. The proposed solution requires multiple RF chains and Channel State Information (CSI) at the relay. A similar architecture is considered in \cite{song2019efficient, fozooni2018hybrid, cai2019robust}. Nevertheless, the channel from the BS to the SR is highly sparse at mmWave bands \cite{xu2019secure}, which restricts the number of spatial streams that can be transmitted. In fact, in practical deployments, the BS and the SR are mounted in over-elevated positions, resulting in a Line-of-Sight (LoS) dominant channel condition, which allows for a single spatial stream transmission. 
This paper proposes a tri-sectoral Advanced SR (ASR), composed of an antenna array oriented toward the BS and two antenna arrays oriented toward two separated serving areas, each having a field of view of 120 deg, as depicted in Fig. \ref{subfig:ASR}. The two arrays serving the UEs can operate in time or frequency division duplexing. The ASR is more than just two SR as the ASR multi-user precoder is optimized to maximize the number of served UEs in the coverage area with constraints on the time-frequency employed resources and on the minimum per-UE rate requirement. The proposed solution is evaluated numerically against the conventional two-panel SR, for an equal number of antennas, showing a significant gain in the achieved cumulative Spectral Efficiency (SE) and the number of served UEs. In particular, the ASR doubles the SE and the number of served UEs with respect to the conventional SR at medium-to-low UE density in space. Moreover, with a suitable trade-off between the maximization of served UEs and the minimization of the employed slots, the gain is even more evident.

\textit{Organization}: The paper is organized as follows: Section \ref{Sec:SystemModel} outlines the system model, Section \ref{sec:ASR} describes the ASR design, with multi-user precoder optimization. 
Numerical results are in Section \ref{sec:simulation} while Section \ref{sec:conclusion} concludes the paper.

\textit{Notation}: Bold upper- and lower-case letters describe matrices and column vectors. Matrix transposition and hermitian are indicated respectively as $\mathbf{A}^{\mathrm{T}}$ and $\mathbf{A}^{\mathrm{H}}$. $\odot$ is the element-wise product between matrices, $\mathbf{I}_n$ is the identity matrix of size $n$. With  $\mathbf{a}\sim\mathcal{CN}(\boldsymbol{\mu},\mathbf{C})$ we denote a multi-variate circularly complex Gaussian random variable $\mathbf{a}$ with mean $\boldsymbol{\mu}$ and covariance $\mathbf{C}$. $\mathbf{1}_N$ denotes an all-ones column vector of size $N$, $\mathbb{E}[\cdot]$ is the expectation operator, while $\mathbb{R}$, $\mathbb{C}$ and $\mathbb{B}$ stand for the set of real complex, and Boolean numbers, respectively. $\delta_{n}$ is the Kronecker delta.

\section{System Model} \label{Sec:SystemModel}
Let us consider an Orthogonal Frequency Division Multiplexing (OFDM) system with $N$ subchannels, where a set $\mathcal{K}$ (cardinality $K$) of User Equipments (UEs) is connected to a BS through an ASR, as depicted in Fig. \ref{subfig:ASR} (the direct link between UEs and BS can be blocked). The $\mathcal{K}$ UEs are uniformly and randomly distributed in the service areas $\mathcal{A}_1$ and $\mathcal{A}_2$ of the ASR panels $P_1$ and $P_2$, respectively, while ASR panel $P_0$ is connected to the BS. Without loss of generality, both BS and ASR are equipped with Uniform Linear Arrays (ULA) of $M_b$ and $M_r$ elements, respectively, while UEs are equipped with a single antenna each.  
\begin{figure}[t]  
 	\centering
	\includegraphics[width=1\columnwidth]{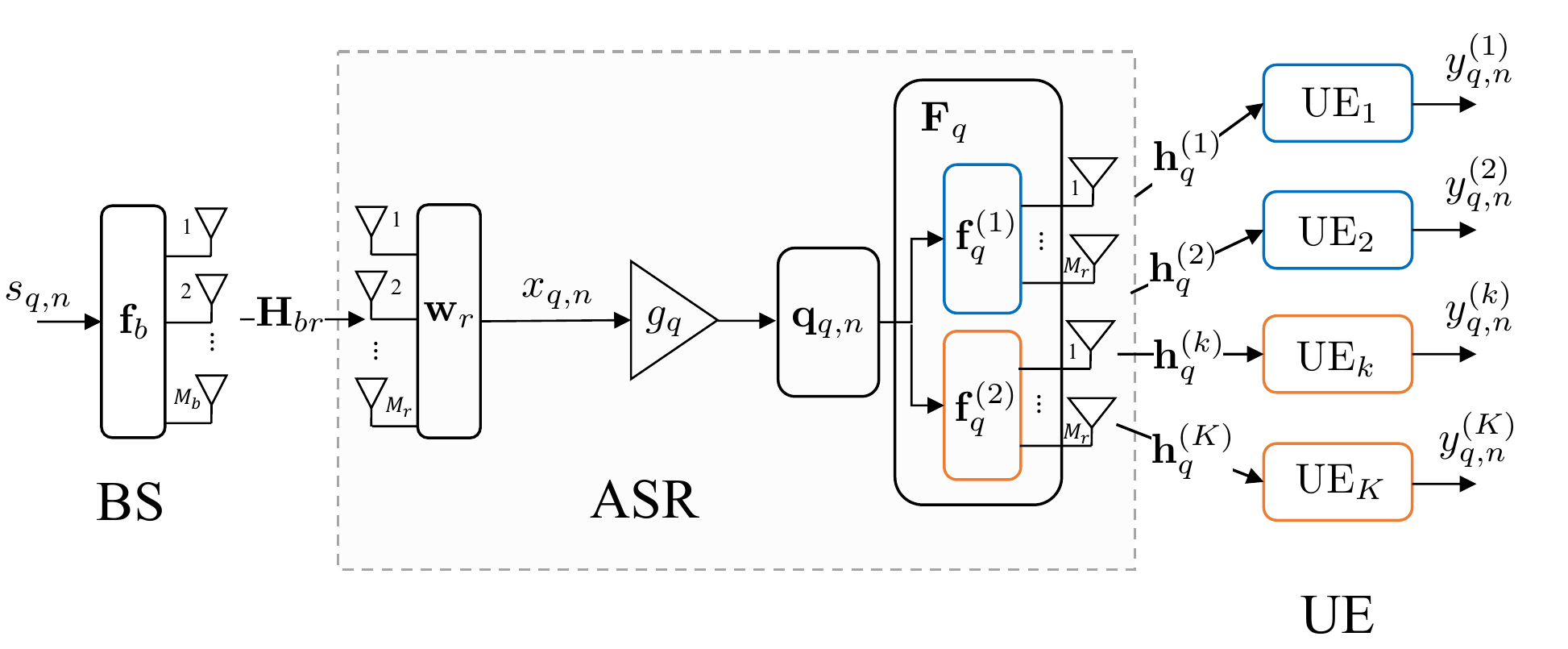} 
    \caption{Block diagram of the ASR-aided DL system model.}
    \label{fig:FrameworkAFRelay}
\end{figure}
In the $q$-th time slot, ruled by the BS, $K_q\leq K$ UEs (in the set $\mathcal{K}_q\subseteq \mathcal{K}$) shall be served through the ASR. On the BS-ASR link (panel $P_0$ in Fig. \ref{subfig:ASR}), the $k$-th UE, $k \in \mathcal{K}_q$, is assigned by the BS with the set of $\mathcal{N}^{(k)}_q\subseteq \{1,...,N\}$ subchannels (of cardinality $|\mathcal{N}^{(k)}_q|=N^{(k)}_q$), such that
\begin{equation}\label{eq:disjoint}
    \mathcal{N}^{(k)}_q \cup \mathcal{N}^{(\ell)}_q = \emptyset\,\, \text{for}\, k\neq \ell
\end{equation}
and 
\begin{equation}\label{eq:disjoint2}
    \sum_{k\in\mathcal{K}_q} N^{(k)}_q \leq N
\end{equation}
where \eqref{eq:disjoint} and \eqref{eq:disjoint2} denote the frequency multiplexing of the $K_q$ UEs as for OFDM.
On the $n$-th subcarrier, the symbol to be transmitted is $s_{q,n}$, uncorrelated between subcarriers, i.e., $\mathbb{E}\left[s_{q,n}s^*_{q,m}\right] = \sigma_s^2\delta_{n-m}$. 
The signal received at the ASR antenna $P_0$ can be expressed as
\begin{equation}\label{eq:rxSigASR}
    x_{q,n} = \mathbf{w}_r^\mathrm{H} \mathbf{H}_{br} \mathbf{f}_b \, s_{q,n} + \mathbf{w}_r^\mathrm{H} \mathbf{n}_{q,n} = \beta \, s_{q,n} + \tilde{n}_{q,n}
\end{equation}
where: $\mathbf{H}_{br} \in \mathbb{C}^{M_r\times M_b}$ denotes the channel matrix between the BS and ASR, $\mathbf{f}_b \in \mathbb{C}^{M_b\times 1}$ and $\mathbf{w}_r \in \mathbb{C}^{M_r\times 1}$ are the frequency-independent Tx and Rx beamforming vectors at the BS and the ASR, respectively, $\mathbf{n}_{q,n} \in \mathbb{C}^{M_r \times 1}$ is the spatial noise vector across ASR antennas, uncorrelated in space, time and frequency, i.e., $\mathbb{E}[\mathbf{n}_{q,n}\mathbf{n}^\mathrm{H}_{\ell,m}] =\sigma_n^2 \mathbf{I}_{M_r} \delta_{q-\ell} \delta_{n-m}$. Notice that $\mathbf{f}_b$ and $\mathbf{w}_r$ are fixed and can be computed during deployment of the BS and ASR, due to the static BS-$P_0$ link, thus the channel amplitude $\beta$ is constant. 

The signal $x_{q,n}$ in \eqref{eq:rxSigASR} is divided into two streams, one for each ASR panel ($P_1$ or $P_2$), such that
\begin{equation}\label{eq:SignalSplit}
    \mathbf{x}_{q,n} = g_q \,\mathbf{q}_{q,n}  x_{q,n} 
\end{equation}
where $\mathbf{q}_{q,n} \in \mathbb{B}^{2\times 1}$ is a selection vector that allocates the signal $x_{q,n}$ to either $P_1$ or $P_2$, depending on the scheduling strategy and ASR operating mode while $g_q$ is the amplification factor. The signal $\mathbf{x}_{q,n}$ is then beam-forwarded to the UEs over $P_1$ or $P_2$. Hence, the signal received by the $k$-th UE can be expressed as
\begin{equation}\label{eq:kthRxSig}
    \begin{split}
        y^{(k)}_{q,n} & = \mathbf{h}^{(k)
        }_{q} \mathbf{F}_q \,\mathbf{x}_{q,n} + z^{(k)}_{q,n}  \\ 
        & = g_q \beta \mathbf{h}^{(k)
        }_{q} \mathbf{F}_q \mathbf{q}_{q,n} s_{q,n} + g_q \mathbf{h}^{(k)
        }_{q} \mathbf{F}_q \mathbf{q}_{q,n} \tilde{n}_{q,n} +  z^{(k)}_{q,n}
    \end{split}
\end{equation}
where $\mathbf{h}^{(k)}_q \in \mathbb{C}^{1 \times 2M_r}$ is the channel vector between the ASR antennas and the $k$-th UE, $\mathbf{F}_q\in \mathbb{C}^{2M_r \times 2}$ is the frequency-independent precoding matrix applied at the ASR, and $z^{(k)}_{q,n}$ is the additive Gaussian nose at the UE, such that $\mathbb{E}[z^{(k)}_{q,n}z^{(k),*}_{\ell,m}] =\sigma_z^2 \delta_{q-\ell} \delta_{n-m}$. 
\subsection{Channel Model}

The mmWave channels have limited paths due to the harsh propagation conditions \cite{rappaport2019wireless}. Hence, we consider the spatially sparse MIMO channel model \cite{mizmizi2021channel, alkhateeb2014channel}. The channel matrix can be expressed as the sum of $P$ paths (usually $P\leq 2$), such that
%
%
\begin{equation} \label{eq:channel_model}
    \mathbf{H} = \sum_{p = 1}^P \alpha_p \varrho_r(\theta_p) \varrho_t(\psi_p) \mathbf{a}_r(\theta_p)
    \mathbf{a}^H_t(\psi_p)
\end{equation}
where $\alpha_p$ is the complex channel gain of the $p$-th path, which accounts for the path-loss and phase effects (e.g., Doppler shift), $\mathbf{a}_t(\psi)$ and $\mathbf{a}_r(\theta)$ denote the Tx and Rx response vectors, respectively, function of the Angle of Departure (AoD) $\psi$ and Angle of Arrival (AoA) $\theta$, while $\varrho(\cdot)$ is the element pattern response. 
The array response vector for a half-wavelength-spaced $M$-element ULA at center bandwidth is:
\begin{equation} \label{eq:array_response_ula}
    \mathbf{a}(\theta) = \left[1, e^{j \pi \text{sin}\theta}, \dots, e^{j \pi (M - 1) \text{sin}\theta} \right]^\mathrm{T}.
\end{equation}
%
%

\section{Design of Advanced SR} \label{sec:ASR}
In conventional SR, the signal is not converted to the digital domain, thus it does not introduce any processing delay. An SR has two antenna arrays, one directed toward the BS, and the other toward the service area. Each antenna can effectively serve UEs within the azimuth range $[\varphi-60, \varphi+60]$ deg \cite{Flamini9720231}. Additionally, to guarantee robustness against self-interference, the angle difference between the two antennas must be $\geq 120$ deg \cite{askar2021interference}. All these constraints limit the covered area of the SR. 
The proposed ASR has instead three panels, each covering a sector of $120$ deg. In the following, we detail the multi-user codebook proposed for the ASR, and then we describe the multi-user precoder optimization. 

\subsection{Codebook Design}\label{subsec:codebook}
The antenna arrays of the ASR are fully analog, implying that the precoding matrix $\mathbf{F}_q$ in \eqref{eq:rxSigASR} can be expressed as
\begin{equation}
    \mathbf{F}_q = \begin{bmatrix}
        \mathbf{f}_q^{(1)}  & \mathbf{0}\\
        \mathbf{0}      & \mathbf{f}_q^{(2)}
    \end{bmatrix}
\end{equation}
where $\mathbf{0}$ is an all zeros vector of size $M_r$, and $\mathbf{f}_q^{(p)} \in \mathbb{C}^{M_r\times 1}$ is the analog precoding vector for the $p$-th panel, whose elements are characterized by $|[\mathbf{f}_q^{(p)}]_i|=1$ for $p=1,2$ and $i=1,...,M_r$. 
Here, we consider a multi-level codebook to constrain the analog precoding vector $\mathbf{f}^{(p)}_q$. Since each antenna array serves a $120$ deg slice of the coverage area, we define a grid of $D$ equi-spaced angles such that
\begin{equation}\label{eq:MainSet}
\Theta = \left\{-60 \leq \theta_d \leq 60\, \text{deg} \; \big\lvert \;  |\theta_d - \theta_{d+1}| = \delta_\theta \right\}
\end{equation}
where $\delta_\theta$ is the grid resolution. The proposed codebook $\mathcal{F}$ has $L$ levels. Let us divide the set $\Theta$ into $L$ non overlapping subsets $\Theta_v$, i.e., $\Theta_u \cap \Theta_v = \emptyset$ for $u \neq v$, such that 
\begin{equation}
    \bigcup\limits_{v=1}^L \Theta_v = \Theta
\end{equation}

The $u$-th entry of the codebook at level $\ell$ can be expressed as
\begin{equation}\label{eq:codebookDef}
    \mathbf{f}_u^{(\ell)} = \left(\mathbf{A} \mathbf{A}^\mathrm{H}\right)^{-1} \mathbf{A}^\mathrm{H} \mathbf{b}_u^{(\ell)}(\theta)
\end{equation}
where $\mathbf{A} = \left[\mathbf{a}(\theta_1), \cdots, \mathbf{a}(\theta_D)\right]^\mathrm{T}$ is a dictionary of $D$ beams and $\mathbf{b}_u^{(\ell)}(\theta) =  \left[b^{(\ell)}_u(\theta_1), \cdots, b^{(\ell)}_u(\theta_D)\right]^\mathrm{T}\in\mathbb{B}^{D\times 1}$ is a Boolean vector defined as
\begin{equation}\label{eq:bVector}
    b^{(\ell)}_u(\theta_d) = 
    \begin{cases}
	1, & \mathrm{if} \,\,  \theta_d \in \Theta_u^{(\ell)} \\
	0 & \mathrm{otherwise}
    \end{cases}
\end{equation}
in which the set $\Theta_u^{(\ell)}$ is the union of $\ell$ subsets $\Theta_v$ and the index $u$ denotes one of the possible permutations. For the $\ell$-th level, the number of permutations, thus the number of beams, is:
\begin{equation}
    \binom{L}{\ell} = \frac{\ell!}{L!(L-\ell)!}
\end{equation}
\begin{figure}[!b]  
	\centering
	{\includegraphics[width=0.4\textwidth]{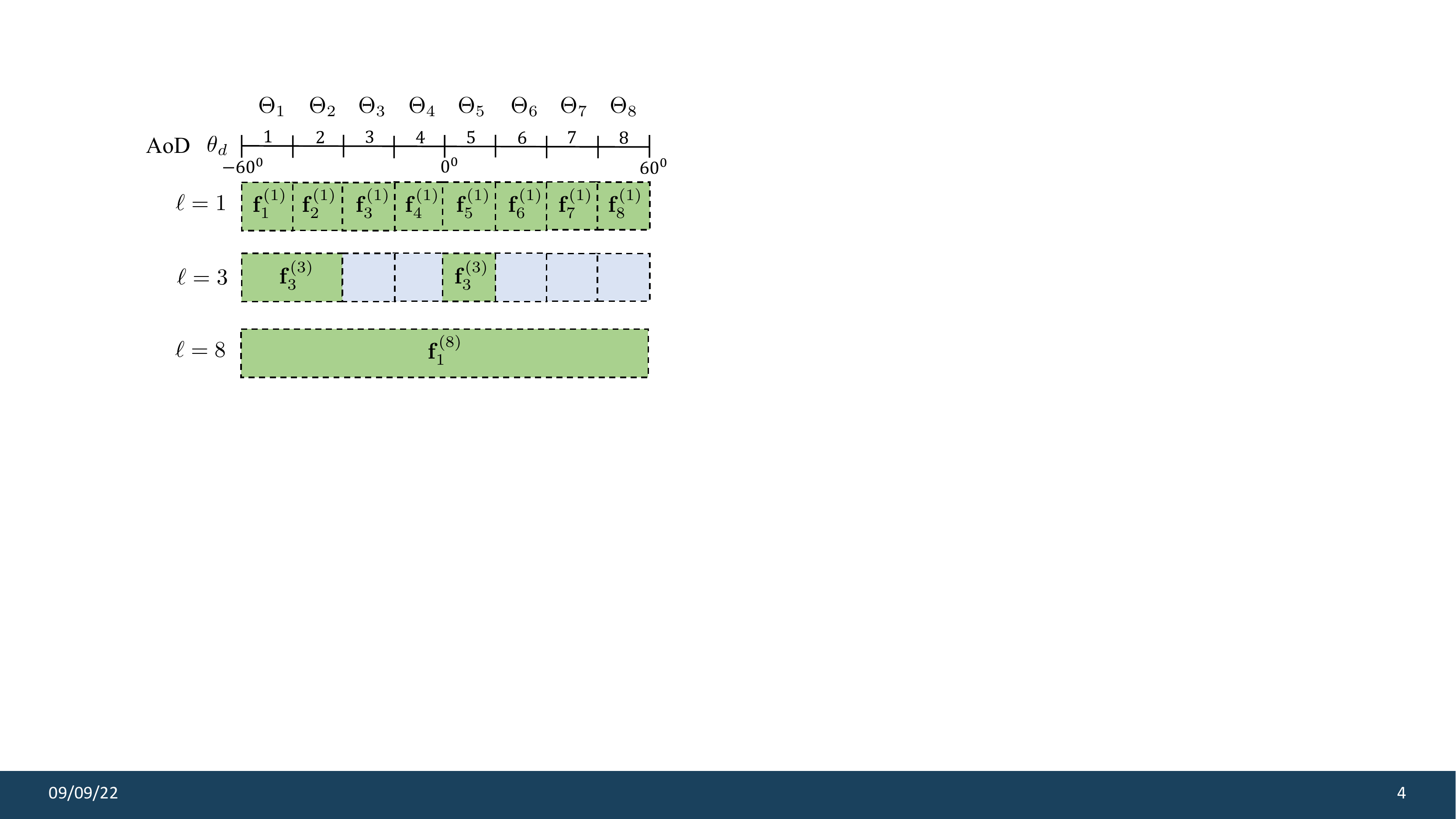}} 
    \caption{Example of three levels for a  codebook with $L=8$, subsets $\Theta_v$ of the same size. Subsets highlighted in green represent active angles spanned by the beamformer. The number of beams per level is 8, 28, 56, 70, 56, 28, 8, 1. }
    \label{fig:codebookStructure}
\end{figure}

\begin{figure*}[!tb]  
	\centering
	\subfloat[Codebook level $\ell=1$]{\includegraphics[width=0.48\columnwidth]{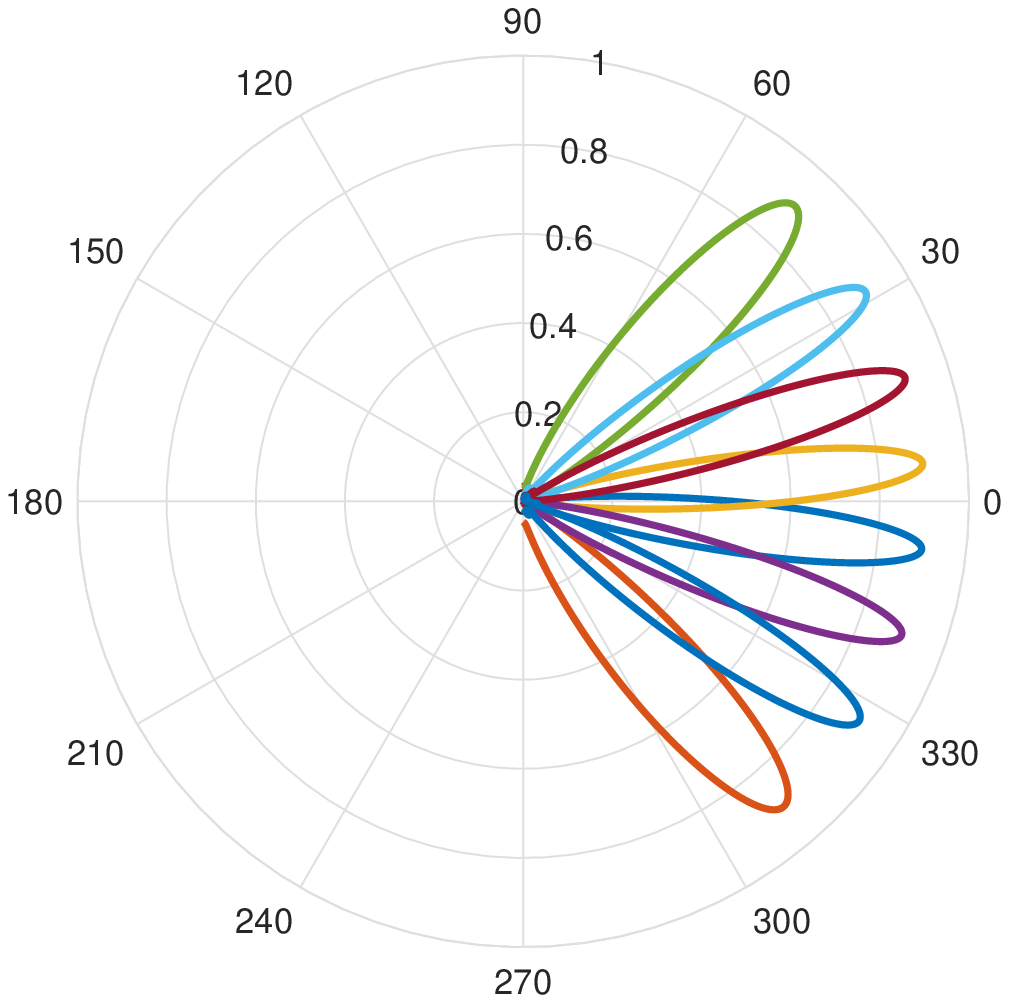}} \,
	\subfloat[Codebook level $\ell=3$]{\includegraphics[width=0.48\columnwidth]{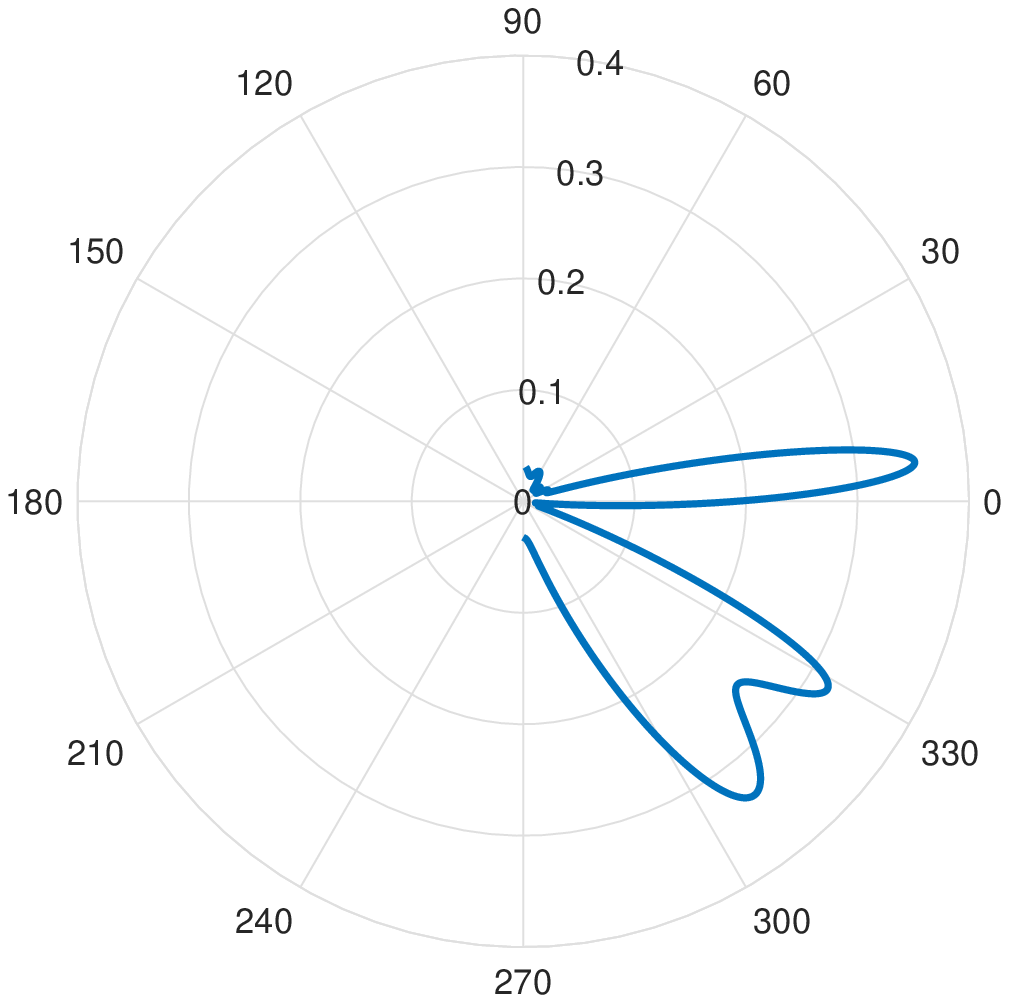}} \,
	\subfloat[Codebook level $\ell=8$]{\includegraphics[width=0.48\columnwidth]{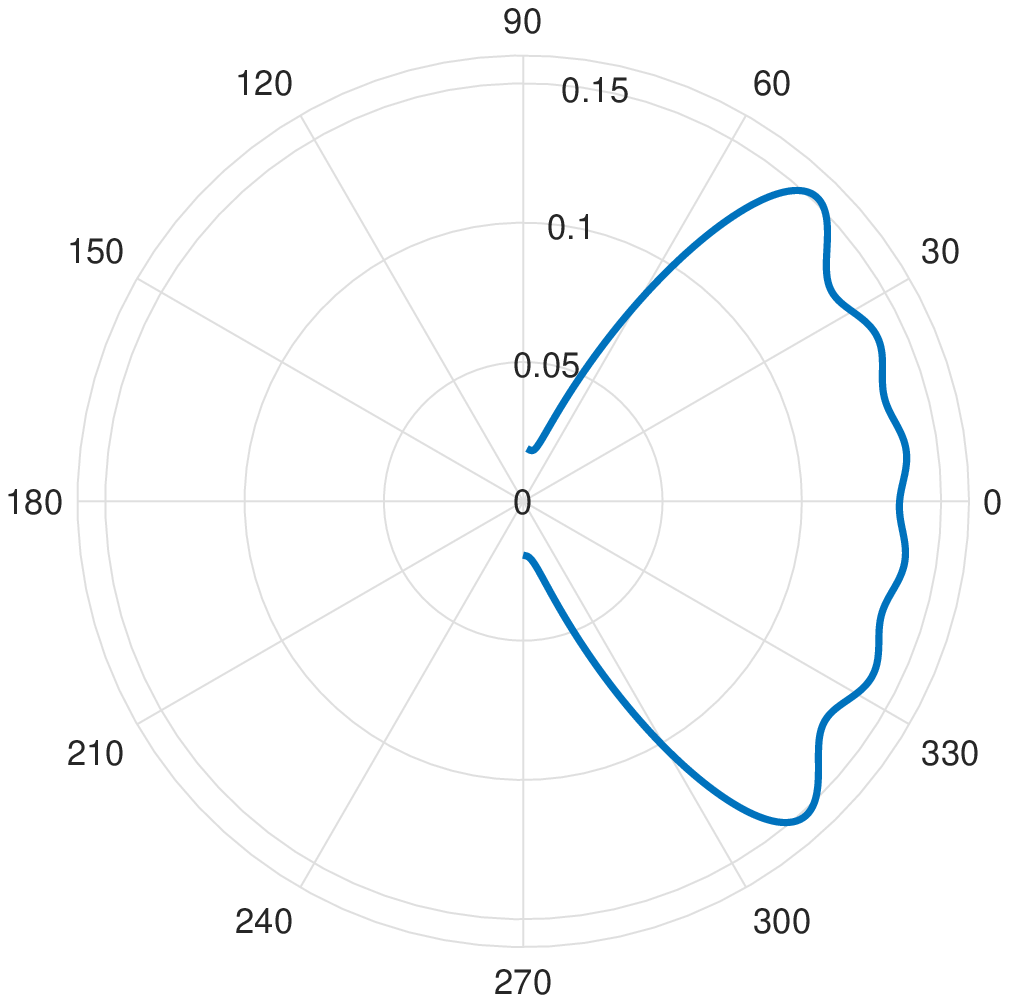}}\,
    \caption{Exemplary codebook beams corresponding to Fig. \ref{fig:codebookStructure}.}
    \label{fig:beampattern}
\end{figure*}
An example of beamformers provided by the proposed codebook is reported in Figs. \ref{fig:codebookStructure} and \ref{fig:beampattern}, for $L=8$, subsets all of the same size. For level $\ell=1$ of the codebook, the number of permutations is obviously 8, thus we have 8 beams pointing in the directions dictated by the subsets $\Theta_v$, $v=1,...,8$. For $\ell=3$, the beam identified by Fig. \ref{fig:codebookStructure} has two main pointing directions ($\Theta^{(3)}_u = [-60, -30) \cup [0, 15)$ deg) and variable beamwidths. Finally, $\ell=8$ implies one possible beam, approximately omnidirectional. Notice that, since we consider phase-only beamforming, the beamforming gain decreases for increasing $\ell$. 

%

\subsection{Multi-User Precoder Optimization} \label{sec:prob_formulation}

The precoder schedules beam and served UEs for the multi-user scenario. Following the 3GPP architecture, the available time-frequency resources are split into $Q$ time slots, comprising 14 OFDM symbols. Without loss of generality, each UE is assigned a single sub-channel and time slot. This assumption is equivalent to assuming that all users have the same Quality-of-service (QoS) requirements. Moreover, we assume that, in a single time slot, only one beam can be employed according to the implementation constraints of ASR. The goal of precoder design is to obtain $\mathbf{F}_q$, $q=1,...,Q$, maximizing the number of served UEs in a given time interval, represented by the $Q$ time slots, ensuring a minimum rate for each served UE.
Specifically, the individual user rate can be defined as
\begin{equation}\label{eq:su_rate}
    \eta_q^{(k)}(\mathbf{F}_q) = \log_2\left(1+\gamma^{(k)}_q(\mathbf{F}_q)\right)
\end{equation}
where $\gamma_q^{(k)}(\mathbf{F}_q)$ is the $k$-th UE signal-to-noise ratio (SNR), that depends on $\mathbf{F}_q$ as based on system model \eqref{eq:kthRxSig}
\begin{equation}\label{eq:su_snr}
\begin{split}
    \gamma_{q}^{(k)}(\mathbf{F}_q) & = \sum_{n\in\mathcal{N}_q^{(k)}}\frac{ \mathbb{E}\left[\lvert g_q \beta \mathbf{h}^{(k)
        }_{q} \mathbf{F}_q \mathbf{q}_{q,n} s_{q,n} \rvert^2\right]}{\mathbb{E}\left[\lvert g_q \mathbf{h}^{(k)
        }_{q} \mathbf{F}_q \mathbf{q}_{q,n} \tilde{n}_{q,n} \rvert^2\right] + \sigma^2_z } = \\
        & \overset{(a)}{=}\frac{ g^2_q |\beta|^2 \sigma^2_s\,\big\lvert \mathbf{h}^{(k)
        }_{q} \mathbf{f}^{(p)}_q \big\rvert^2}{g^2_q \sigma^2_n M_r \big\lvert \mathbf{h}^{(k)
        }_{q} \mathbf{f}^{(p)}_q \big\rvert^2 + \sigma^2_z }.
\end{split}
\end{equation}
where we made the assumption that the $k$-th UE is served by either $P_1$ or $P_2$ ASR panel, namely all the subcarriers in $\mathcal{N}_q^{(k)}$ are scheduled to the same panel (equality $(a)$). By assuming that in a single time slot, only one beam per panel is employed, the SNR becomes 
\begin{equation}\label{eq:su_snr_singlebeam}
    \gamma_{b}^{(k)} = \frac{ g^2_q |\beta|^2 \sigma^2_s\,\big\lvert \mathbf{h}^{(k)
        }_{q} \mathbf{f}^{(p)}_b \big\rvert^2}{g^2_q \sigma^2_n M_r \big\lvert \mathbf{h}^{(k)
        }_{q} \mathbf{f}^{(p)}_b \big\rvert^2 + \sigma^2_z }
\end{equation}
where we dropped the index $q$ on the precoder, substituting it with the index of the beam $b=1,...,B$, for the total number of beams 
\begin{equation}\label{eq:num_beams}
    B = \sum_{\ell=1}^L \binom{L}{\ell} = (2^L - 1).
\end{equation}
The multi-user precoder optimization problem can be formulated as the maximization of the number of served UEs within $Q$ time slots ensuring a single user rate QoS requirement. We make the following assumptions: \textit{(i)} channels $\mathbf{h}^{(k)}_{q}$ are known at the ASR, $\forall q,k$, (e.g., via suitable estimation) and \textit{(ii)} each UE is associated with only one of the panels according to the served areas (Fig. \ref{subfig:ASR}). This latter is equivalent to asserting that $\mathbf{q}_{q,n}$ is known $\forall q,n$.
To tackle the problem, we define the association variables
\begin{equation}\label{eq:association_var_1}
   a_{k,b} =  \begin{cases} 1,\,\, \text{if the $k$-th user is served by the $b$-th beam},\\
   0,\,\,\text{otherwise}.
    \end{cases}
\end{equation}
stacked into vector 
$\mathbf{a}_k = [a_{k,1},a_{k,2},...,a_{k,2B}]^\mathrm{T}\in \mathbb{B}^{2B\times1}$ and 
\begin{equation}\label{eq:association_var_2}
   {z}_b^{(p)} =  \begin{cases} 1,\,\, \text{if the $b$-th beam refers to the $p$-th panel},\\
   0,\,\,\text{otherwise}.
    \end{cases}
\end{equation}
stacked in vector $\mathbf{z}^{(p)} = [z_1^{(p)}, z_2^{(p)}, ...., z_{2B}^{(p)}]\in\mathbb{B}^{1\times 2B}$. In \eqref{eq:association_var_1} and \eqref{eq:association_var_2} we consider that the total number of beams that can be used to serve the $K$ UEs (served by both $P_1$ and $P_2$ panels) are $2B$, namely $B$ for panel $P_1$ and $B$ for panel $P_2$.
The optimization problem is formulated as: 
\begin{subequations}
\begin{alignat}{2} 
&\underset{\mathbf{a}_1,...\mathbf{a}_K}{\text{max}}  &\quad& \sum_{k=1}^K \mathbf{a}_k^\mathrm{T}\mathbf{1}_{2B}\label{eq:optProb}\\
&\text{s.t.} &      & \mathbf{a}_k^\mathrm{T}\mathbf{1}_{2B} \leq 1,\,\,\,\forall k,\label{eq:prob1_constraint1}\\
&                  &      & \sum_{k=1}^K \mathbf{a}_k^\mathrm{T}\mathbf{1}_{2B} \leq N,\label{eq:prob1_constraint2}\\
&                  &      & \log_2\left(1+ a_{k,b}\,\gamma^{(k)}_b\right) \geq  a_{k,b}\,\overline{\eta},\,\,\, \forall k, \forall b,\label{eq:prob1_constraint3}\\
&                  &      & \norm{\sum_{k=1}^K \mathbf{a}_k \odot \mathbf{z}^{(p)}}_0 \leq Q,\,\,\,p = 1,2,\label{eq:prob1_constraint4}\\
&                  &      & a_{k,b} \in \{0,1\},\label{eq:prob1_constraint5}
\end{alignat}
\end{subequations}
where the objective function in \eqref{eq:optProb} refers to the maximization of the total number of served UEs. The constraint in \eqref{eq:prob1_constraint1} forces each UE to be assigned to a single beam and, equivalently to a single time slot, while the maximum number of served UEs is upper bounded by the total number of sub-channels $N$ (constraint \eqref{eq:prob1_constraint2}). The QoS requirement is defined by the constraint in \eqref{eq:prob1_constraint3}, where $\gamma^{(k)}_b$ is the SNR experienced by the $k$-th UE when served by the $b$-th beam as in \eqref{eq:su_snr_singlebeam}, and $\overline{\eta}$ is the user rate threshold. The constraint \eqref{eq:prob1_constraint4} specifies the upper bound for the number of scheduled beams for each panel and, equivalently, the scheduled time slots for each panel. 
The constraint ensures that the maximization of the served UEs is performed by achieving a solution that guarantees the constraint \eqref{eq:prob1_constraint4} to be at most $Q$-sparse. However, the L$_0$–norm operator is non-convex and discontinuous. Therefore, we reformulate the optimization problem \eqref{eq:optProb} as
\begin{subequations}
\begin{alignat}{2} 
&\underset{\mathbf{a}_1,...\mathbf{a}_K, \boldsymbol{\delta}}{\text{max}}  &\quad& \sum_{k=1}^K \mathbf{a}_k^\mathrm{T}\mathbf{1}_{2B} - \lambda \sum_{b = 1}^B \delta_b \label{eq:optProb2}\\
&\text{s.t.} &      & \mathbf{a}_k^\mathrm{T}\mathbf{1}_{2B} \leq 1,\,\,\,\forall k,\label{eq:prob2_constraint1}\\
&                  &      & \sum_{k=1}^K \mathbf{a}_k^T\mathbf{1}_{2B} \leq N\label{eq:prob2_constraint2}\\
&                  &      & \log_2\left(1+ a_{k,b}\,\gamma^{(k)}_b\right) \geq  a_{k,b}\,\overline{\eta},\,\,\, \forall k, \forall b,\label{eq:prob2_constraint3}\\
&                  &      &  \delta_b \leq \sum_{k=1}^K a_{k,b}  {z}_{b}^{(p)} \leq K\delta_b,\,\,\,\forall b,\,\,p = 1,2,\label{eq:prob2_constraint4}\\
&                  &      & \sum_{b = 1}^B \delta_b {z}_{b}^{(p)}  \leq Q,\,\,\,p = 1,2,\label{eq:prob2_constraint5}\\
&                  &      &a_{k,b} \in \{0,1\},\label{eq:prob2_constraint6}\\
&                  &      &\delta_{b} \in \{0,1\},\label{eq:prob2_constraint7}
\end{alignat}
\end{subequations}
where $\boldsymbol{\delta} \in \mathbb{B}^{2B\times 1}$ is a vector of auxiliary binary variables, $\lambda$ is the regularization parameter that defines the trade-off between the number of scheduled beams (i.e., slots) for the two panels and the number of served UEs. By setting $\lambda >0$, the objective function in \eqref{eq:optProb2} becomes a trade-off between the number of served users and the number of needed time slots. On the other hand, if $\lambda\leq 0$, \eqref{eq:optProb2} aims only to maximize the number of served UEs by satisfying the constraints in \eqref{eq:prob2_constraint4} and \eqref{eq:prob2_constraint5}. The optimization problem \eqref{eq:optProb2} is a relaxation of \eqref{eq:optProb}, taking the form of an Integer Linear Program (ILP). ILP in \eqref{eq:optProb2} is NP-hard, but, since it involves binary optimization variables, it can be cast as a \textit{knapsack} problem, to be solved in pseudo-polynomial time using dynamic programming \cite{papadimitriou1981complexity}.

\section{Numerical Simulations} \label{sec:simulation}

The performance of the proposed tri-sectoral ASR is numerically evaluated and compared with the conventional two-panel SR counterpart as a benchmark. The set of simulation parameters is in Table \ref{ParameterSetups}, which are used in any setup unless otherwise specified. In particular, the $K$ UEs are randomly distributed in the whole service area, namely spanning the 240 deg coverage of the ASR. Moreover, the number of antennas for panel $P_1$ of the conventional SR is $2 M_r=16$, while the tri-sectoral ASR is equipped with panels of $M_r=8$ antennas each ($P_1$ and $P_2$). This choice has been adopted to make the fairest comparison that preserves the hardware complexity of the relays. While ASR implements the multi-user codebook described in Section \ref{subsec:codebook} (all levels), the conventional SR implements only the first level, corresponding to the standard DFT codebook with $2M_r$ beams.

\begin{table}[!b]
\caption{Simulation Parameters}
\centering
\begin{tabular}{lcc}
\toprule 
\textbf{Parameter} & \textbf{Symbol} & \textbf{Value} \\
\toprule
Frequency & $f_0$ & $28$ GHz\\
Tx power (BS) & $\sigma_s^2$ &  $30$ dBm \\ 
Amplification factor (ASR) & $g^2_q$ &  $1/|\beta|^2$ \\
BS antennas & $M_b$ &  64 \\
ASR antennas & $M_r$ &  8 \\
Radius of the serving area & $R$ &  100 [m] \\
BS-ASR distance  & $d_{br}$ &  200 [m] \\
Noise power level (ASR) & $\sigma^2_n$ &  $-85$ dBm \\
Noise power level (UE) & $\sigma^2_z$ &  $-90$ dBm \\
Number of UEs (max) & $K$ & $10,40$ \\
Number of subchannels & $N$ & $8$ \\
Number of time slots & $Q$ & $1,6$ \\
\bottomrule
\end{tabular}
\label{ParameterSetups}
\end{table} 

The performance comparison between SR and ASR is made in terms of cumulative spectral efficiency over the served UEs and time slots, defined as
\begin{equation}
    \xi = \sum_{q=1}^Q \sum_{k\in \mathcal{K}_q}\eta_q^{(k)}
\end{equation}
evaluated separately for SR and ASR when solving problem \eqref{eq:optProb2}. We set the regularization parameter $\lambda = \{-1,1,2\}$. Positive values of $\lambda$ determine a trade-off between the number of served UEs and the number of used time slots as for \eqref{eq:optProb2}. 

\begin{figure}[!t]  
	\centering
\subfloat[][$K=10$]{\includegraphics[width=0.95\columnwidth]{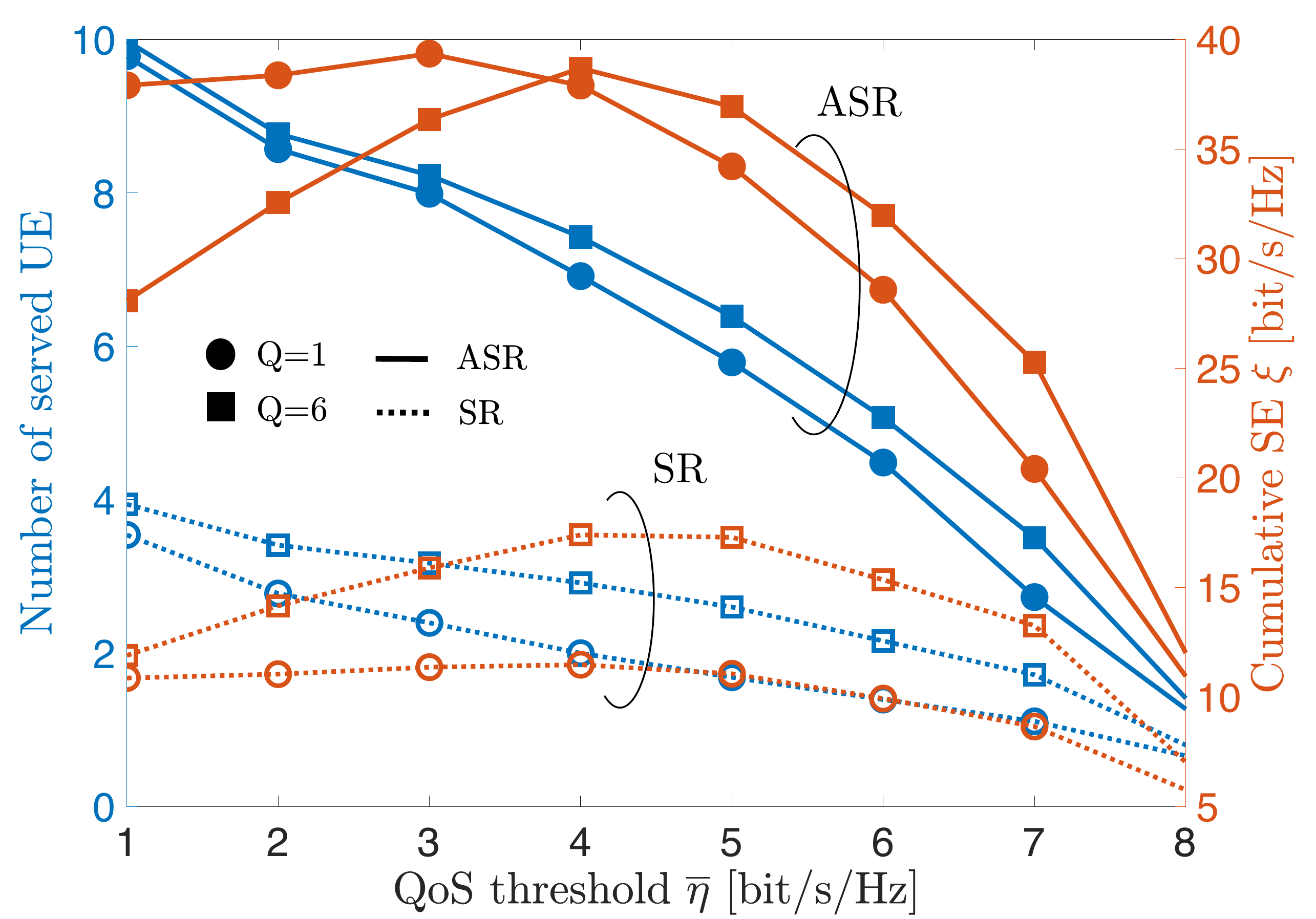}\label{subfig:result1_10}}\\
 \subfloat[][$K=40$]{\includegraphics[width=0.95\columnwidth]{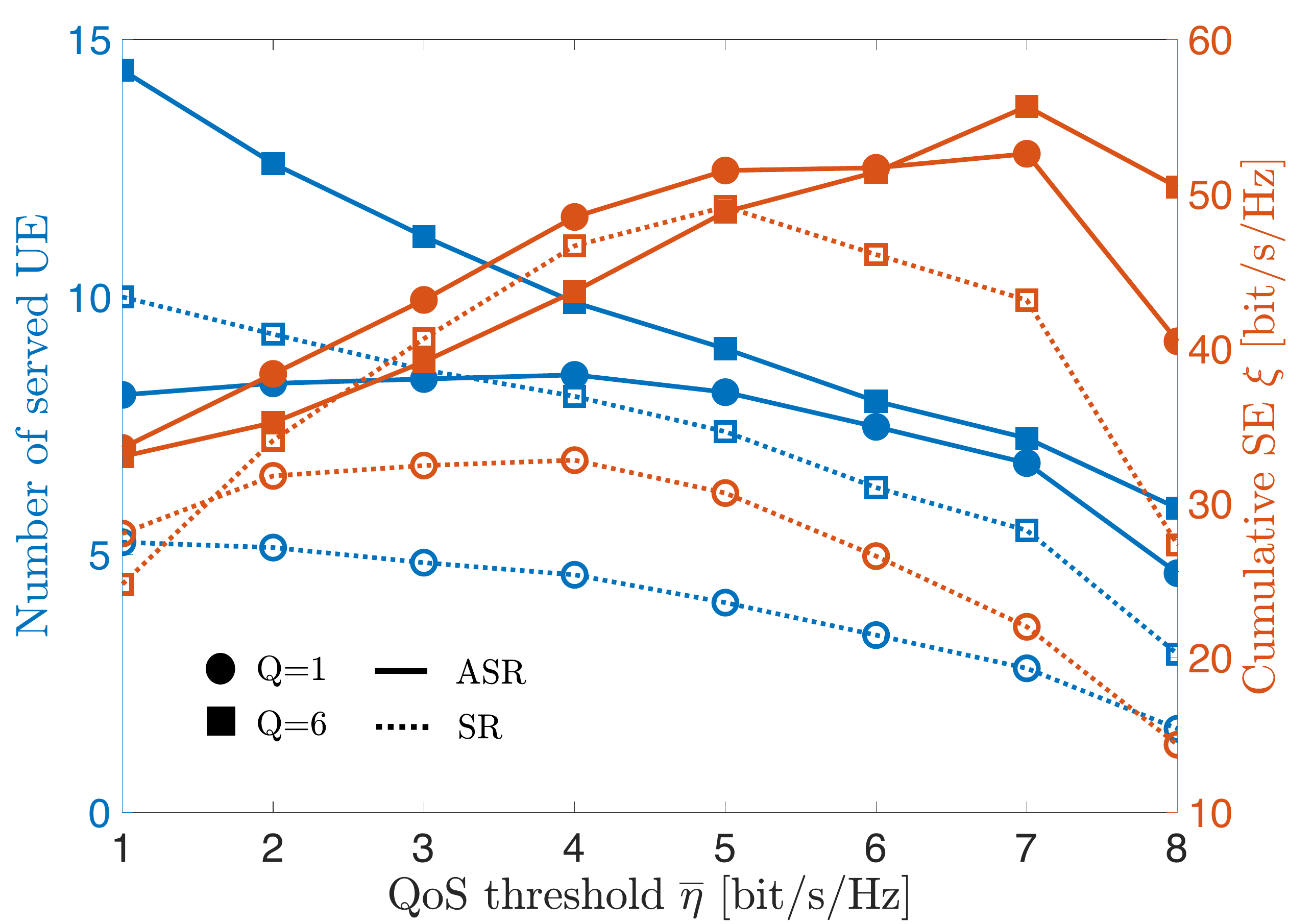}\label{subfig:result1_40}}\\
 \caption{Cumulative SE $\xi$ and number of served UEs varying the QoS threshold $\overline{\eta}$, for both ASR and conventional SR, $Q=1,6$ time slots, $\lambda=-1$ and (\ref{subfig:result1_10}) $K=10$ UEs (max) (\ref{subfig:result1_40}) $K=40$ UEs (max)}
 \label{fig:result1}
\end{figure}

The first result is summarized in Fig. \ref{fig:result1}, comparing the cumulative SE $\xi$ and number of served UEs of ASR and conventional SR varying the QoS threshold $\overline{\eta}$, for $Q=1,6$ time slots and $K=10$ UEs (Fig. \ref{subfig:result1_10}) and $K=40$ UEs (Fig. \ref{subfig:result1_40}), $\lambda=-1$. In both cases, the number of served UEs decreases with the required QoS $\overline{\eta}$. This is expected, as increasing $\overline{\eta}$ implies increasing the required SNR at each UE, thus the directivity and spatial selectivity of the employed beams. The cumulative SE, instead, shows a peak that depends on $K$.
The remarkable benefit of employing ASR is evident, especially at low UE density. For $K=10$, i.e., a medium-to-low UE density, ASR can double the number of served UEs (on average) regardless of the per-UE QoS requirement and the number of employed time slots $Q$. The cumulative SE $\xi$ follows the same trend. The reason for the ASR-to-SR gain is twofold: the ASR covers a doubled area compared to the conventional SR, and can synthesize flexible codebooks to trade between angular coverage and beamforming gain. As the UEs density increases, $K=40$, the ASR-to-SR gain decreases, with a limited benefit in favour of the ASR solution. However, notice that any relaying solution, either ASR or SR, is conceived to bridge the coverage gap of the BS in those regions where most of the UEs are in non-LoS. If the UEs density in these regions gets too high, any AF relay does not perform as expected, pushing for network densification.

\begin{figure}[!t]  
	\centering
    \includegraphics[width=0.88\columnwidth]{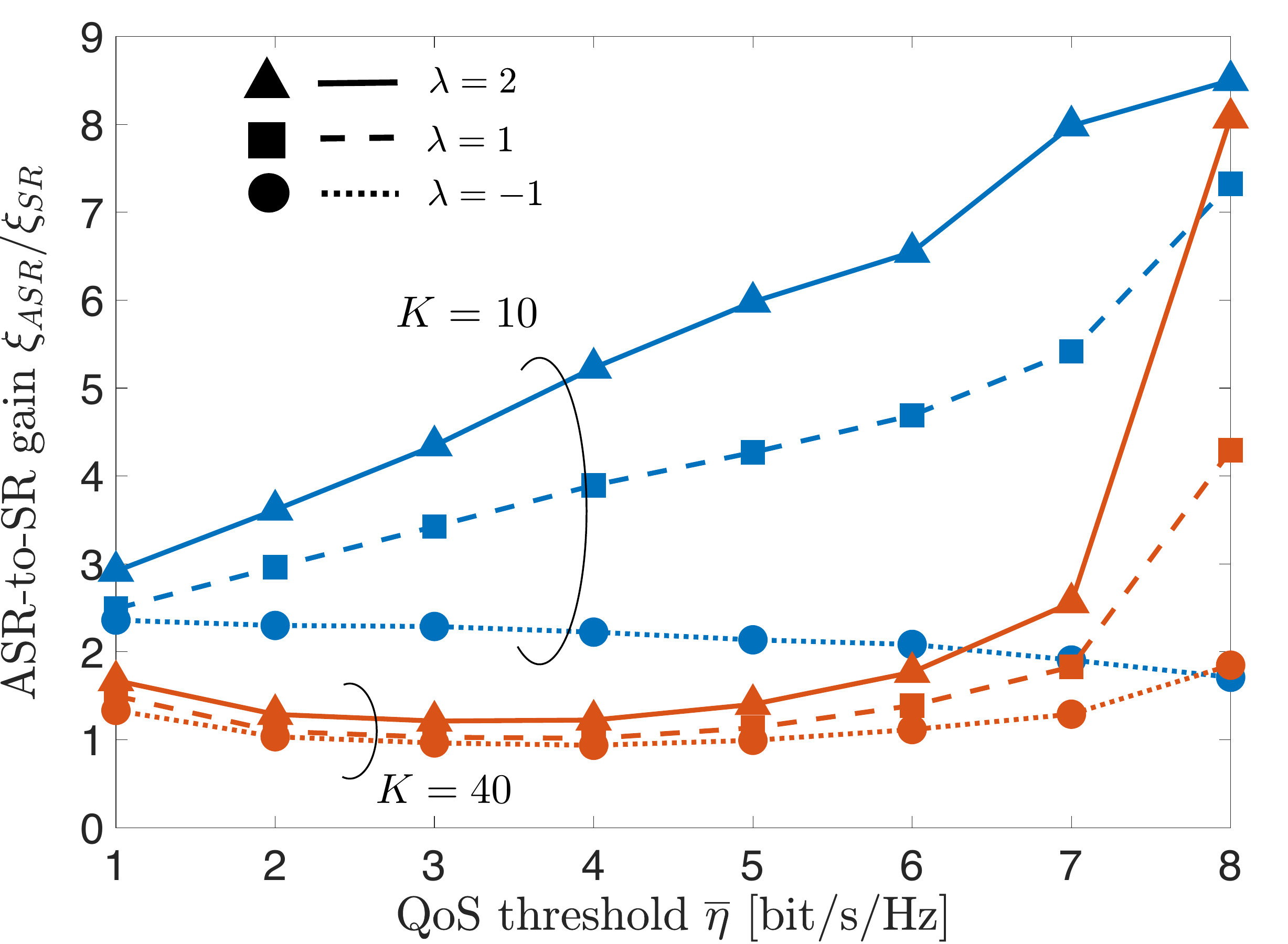}
	\caption{ASR-to-SR gain varying QoS threshold $\overline{\eta}$ for different values of the regularization parameter $\lambda$, $Q=6$ time slots}
	\label{fig:chivslambda}
\end{figure}

The proper tuning of the regularization parameter $\lambda$ offers the possibility of further improving ASR performance. By setting $\lambda>0$ in \eqref{eq:optProb2}, we can balance the need for serving more UEs with the need of using fewer slots. Fig. \ref{fig:chivslambda} shows the trend of the ASR-to-SR gain $\xi_{ASR}/\xi_{SR}$ varying $\overline{\eta}$ for $\lambda=\{-1,1,2\}$, $K=10$ and $K=40$, $Q=6$. Interestingly, for $\lambda>0$, the ASR-to-SR SE gain shows a positive trend with $\overline{\eta}$, in contrast to the case in which $\lambda=-1$. Notice that enforcing a decrease in the number of employed slots means reducing the BS load in terms of spatial resources employed for the BS-$P_0$ link. The results justify the consideration of the proposed tri-panel ASR as an alternative and high-performing solution for range extension in mmWave and sub-THz networks.


\section{Conclusions}\label{sec:conclusion}
The limited range of mmWave and sub-THz networks, envisaged for capacity-demanding 6G services, call for cost-effective range-extender solutions. This paper investigates the use of tri-panel ASR to double the angular coverage of conventional two-panel SR, currently limited to a sector of 120 deg. We propose a multi-user precoder design for ASR to maximize the number of served UEs in a given time interval. Numerical results highlight the benefits of the proposed ASR with respect to conventional SR in terms of cumulative SE gain (an average factor 2) varying the QoS requirement at the UE and the number of UEs in the coverage region.    
%
%
\bibliographystyle{IEEEtran}
\bibliography{reference}

\end{document}